\documentclass[twocolumn,showpacs,floatfix,amsmath,amssymb,prb]{revtex4}
\usepackage{graphicx}
\usepackage{amsmath}
\usepackage{dcolumn}
\usepackage{bm}
\newcommand{\Hamil}{{\cal H}}
\newcommand{\kk}{{\mathbf k}}

\begin{document}

\title{Temperature dependence of polaronic transport through\\
single molecules and quantum dots}

\author{Urban Lundin}
 \email{lundin@physics.uq.edu.au}
\author{Ross H.\ McKenzie}
\affiliation{Department of Physics, University of Queensland, 
             Brisbane Qld 4072, Australia}

\date{\today}

\begin{abstract}
Motivated by recent experiments on electric transport through 
single molecules and quantum dots, 
we investigate a model for transport that allows
for significant coupling between
the electrons and a boson mode isolated on the molecule or dot. 
We focus our attention on the temperature dependent properties 
of the transport. In the Holstein picture
for polaronic transport in molecular crystals 
the temperature dependence of the conductivity exhibits a 
crossover from coherent (band) to incoherent (hopping) transport.
Here, the temperature dependence of the
differential conductance on resonance does not show 
such a crossover, but is mostly determined by
the lifetime of the resonant level on the molecule or dot. 
\end{abstract}
\maketitle

\section{Introduction}
In recent years there has been a growing interest in electrical transport 
through single molecules~\cite{chen99,park00,bockrath97,joachim00}
and single electronic levels in 
quantum dots.~\cite{schmidt97,fujisawa98,qin01} 
Some molecular devices exhibit 
switching behavior with large on-off ratios~\cite{chen99} 
increasing the motivation to construct 
molecular electronic devices.~\cite{joachim00}
In some cases it has been found that the
transport is quite temperature dependent~\cite{chen99} 
and it has been suggested~\cite{ventra01} that this
is due to the presence of low energy boson modes, such as internal rotations,
which couple strongly to the molecular electronic states, 
and can easily be excited by small 
temperatures.~\cite{hone00,armour01} 
In a similar vein, in double quantum dots it has
been found that there are acoustic phonons which
couple strongly to the electrons~\cite{fujisawa98,qin01}.

Some experimental values for the phonon energy have 
been estimated in various papers. 
In Table~\ref{expt:tab} we give some numbers for reference. 
\begin{table}[h]
\caption{\label{expt:tab}Typical values for parameters taken from experiment.
         $\hbar\omega_0$ is the boson energy and $\Gamma$ is the line width 
         due to coupling to the leads
         (defined below) of the resonant level on the molecule or dot. 
         $I_{max}$ is the maximal current driven through the system.
}
\begin{tabular}{lccc}
System & $\hbar\omega_0$ & $\Gamma$ & $I_{max}$ \\ 
\hline
2 quantum dots\cite{qin01}
  & 40 $\mu$eV   &  0.2 $\mu$eV              & 3   pA \\ 
2 quantum dots\cite{fujisawa98,brandes99}
  & 30 $\mu$eV   & 1 $\mu$eV & 5   pA \\ 
molecule\cite{chen99,ventra01}
  & 3 meV  &                & 1   nA \\ 
C$_{60}$ molecule\cite{park00}
  & 5 meV  &                & 0.1   nA \\ 
\end{tabular}
\end{table}
We see that the boson (usually phonon) frequency in these systems 
is quite small, corresponding to temperatures in
the range 0.5 -- 50 Kelvin.
In addition there was a recent proposal~\cite{armour01} to consider
transport through a quantum dot to
a carbon nanotube cantilever with a resonant frequency
of the order of 100 MHz, corresponding to 
a phonon energy of 0.4 $\mu$eV.
If the electron-phonon coupling is sufficiently large
polaronic transport might be important for these systems. When the electron 
tunnels through it can absorb or emit bosons, thus altering its energy and the 
current. If the temperature is much larger than the 
boson energy, there are many bosons available for absorption and this 
might heavily influence the current.  

In 1959 Holstein~\cite{holstein59} predicted that for a 
periodic one-dimensional molecular crystal with strong
electron-phonon coupling there should be a 
crossover from coherent (band) to incoherent (hopping) 
transport with increasing temperature.
When increasing the temperature the effective bandwidth becomes narrower, this 
gives rise to a decrease in coherent transport. 
In contrast, increasing 
temperature means that more and more phonons are activated and we are in a 
regime where phonon assisted inter-site
tunneling starts to contribute to the 
conductivity. This coherent-incoherent
crossover is believed to have been observed for 
the first time quite recently in single crystals of pentacene.~\cite{schon01} 
One aim of this paper is to see whether a similar crossover 
should be seen in polaronic transport through molecules and quantum dots.
This might be expected because of the mathematical similarity between 
the models for periodic systems and the resonant tunneling case. 
We might expect the tunneling amplitude between the leads and dot to be 
reduced by polaronic effects, thereby reducing the coherent part of the 
conductivity. When increasing the temperature the electrons can tunnel 
with boson assisted transport that enhances the tunneling, possibly leading to 
a crossover behavior. 
There have been many theoretical investigations of 
the effect of phonons on the transport through 
molecules~\cite{seminario00,ventra01,hettler01,boese01,emberly00} 
and quantum 
dots~\cite{kang98,hyldgaard94,wingreen89,konig96,li95,haule}, 
but none of them focuses on the temperature dependence of the current. 
The purpose of this paper is to clarify this aspect of the transport. 
Li \textit{et.\ al}~\cite{li95} included a Hubbard term, but did not 
consider multi-phonon contributions. In a recent paper Emberly and 
Kirczenow~\cite{emberly00} made a thorough analysis of conductance through a 
molecular wire. A set of self-consistent equations where set up and solved to 
give the distribution functions in the leads and molecule, and then 
transmission probabilities were calculated. However, the temperature 
dependence is not addressed in that paper. 

In this paper we perform the analysis for the
simplest possible case, where the electrons interact with 
a single optical boson localized on the dot or molecule.
We anticipate that this is sufficient to illustrate
the main physics in the more complicated case of many bosons,
such as acoustic phonons.
In order to obtain analytical results we have to assume that the 
coupling to the leads is small and the energy level in the dot or molecule is
not too close to the Fermi energy in the leads.~\cite{hewson80,hewson} 
By assuming that the coupling to the leads is small we can calculate the 
effects from the bosons locally on the molecule/dot and then assume that 
the effect on the leads from the bosons is negligible. This enables us to use 
well-known results from mesoscopic transport theory. 
The bosons are possibly most often 
phonons, but since the theory will look identical (assuming linear couping) 
for different types of bosons (phonons, magnons, charge oscillations) we 
will simply refer to ''bosons''. Even in photon assisted tunneling through 
quantum dots side bands have been observed when tuning the photon 
energy.~\cite{oosterkamp97}
In section II we will define the model we use and in section III, we discuss 
the approximations we have to make. Different limits for the current are 
derived in section IV, and in section V we discuss the differential 
conductivity. 

\section{Current through a level coupled to a local boson mode}

\begin{figure}[hb]
\includegraphics*[scale=1.0]{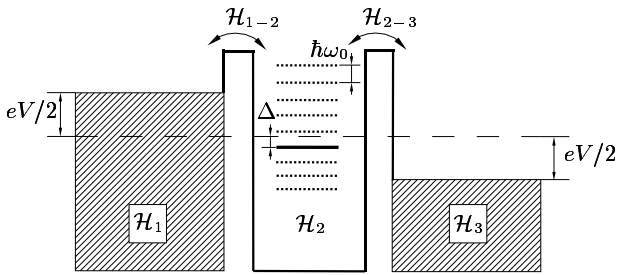}
\caption{Tunneling through a system with one level.
         The dashed lines indicate the bosonic satellites (see text).
         The Fermi energy in the leads is chosen to be zero.
         The electrons has to 
         tunnel through the barriers, and can absorb or emit bosons in the 
         process, corresponding to the lines below and above the central 
         resonance respectively. 
         The Hamiltonian given in Eq.(\ref{hamiltonian:eq}) contains 
         terms describing the different parts of the system.}
\label{descr:fig}
\end{figure}

We consider the simplest possible model Hamiltonian and neglect
the spin degree of freedom and any effects of electron-electron
interactions.
The system we study consists of the individual entities (left lead, molecule 
or quantum dot, and right lead) coupled via tunneling. 
We assume that we are dealing with a resonant tunneling situation, but the 
states in the dot (or molecule or any single level system) couples to some 
boson mode with characteristic frequency $\omega_0$, 
as shown in Fig.~\ref{descr:fig}. 
The Hamiltonian is given by 
\begin{equation}
\Hamil=\Hamil_1+\Hamil_2+\Hamil_3+\Hamil_{1-2}+\Hamil_{2-3}, 
\label{hamiltonian:eq}
\end{equation}
where 
\begin{eqnarray}
&&\Hamil_1+\Hamil_3=
 \sum_{\kk_1}\epsilon_{\kk_1}c^{\dag}_{\kk_1}c_{\kk_1}+
 \sum_{\kk_3}\epsilon_{\kk_3}c^{\dag}_{\kk_3}c_{\kk_3},  \nonumber \\
&&\Hamil_2=
  \epsilon_{0}c^{\dag}_{2}c_{2}+\hbar\omega_0a^{\dag}a+
             Mc^{\dag}_2c_2(a+a^{\dag}),                 \nonumber \\
&&\Hamil_{1-2}=\sum_{\kk_1}t(c^{\dag}_{\kk_1}c_{2}+h.c.),\nonumber \\
&&\Hamil_{2-3}=\sum_{\kk_3}t(c^{\dag}_{\kk_3}c_{2}+h.c.).\nonumber 
\end{eqnarray}
Here $\epsilon_0$ is the energy of the level in the dot/molecule and 
$t$ is the energy associated with hopping 
onto/off the dot. The electronic dispersion in the leads are given by 
$\epsilon_{\kk_1}$ and $\epsilon_{\kk_3}$. $M$ is the coupling to the local 
boson mode with energy $\hbar\omega_0$. We disregard the spin dependence for  
simplicity.

First we make a unitary transformation to diagonalize the Hamiltonian 
$\Hamil_2$. The price we pay for this is that extra operators attach to the 
tunneling term in the Hamiltonian. 
The transformation is 
$\bar{{\cal H}_2}=e^S{\cal H}_2e^{-S}$,
where $S=c^{\dag}c\frac{M}{\hbar\omega_0}(a^{\dag}-a)$.
This gives us
\begin{equation}
\bar{{\cal H}}_2=\hbar\omega_0a^{\dag}a-\Delta c_2^{\dag}c_2,
\label{barH:eq}
\end{equation}
where
\begin{equation}
\Delta=\frac{M^2}{\hbar\omega_0}-\epsilon_0. 
\end{equation}
When the central system is a quantum dot $\epsilon_0$
(and thus $\Delta$) 
can be adjusted by applying a gate voltage. 
After the transformation the tunneling part of the Hamiltonian becomes
\begin{eqnarray}
\bar{\Hamil}_{1-2}=\sum_{\kk_1}t(c^{\dag}_{\kk_1}c_{2}X+h.c.),
   \nonumber \\
\bar{\Hamil}_{2-3}=\sum_{\kk_3}t(c^{\dag}_{\kk_3}c_{2}X+h.c.),
\label{barHt:eq}
\end{eqnarray}
where
\begin{equation}
X=\exp\left[\frac{M}{\hbar\omega_0}
          (a-a^{\dag})\right].
\label{X-op:eq}
\end{equation}
The $X$-factors can be absorbed into a renormalized electron 
creation/annihilation operator in region 2, so that we are left with the 
usual resonant tunneling Hamiltonian except that the Greens function 
for the electrons on the molecule/dot has an additional complication. 
$\langle T_{\tau} c(\tau)c^{\dag}(0)\rangle \rightarrow
 \langle T_{\tau} \bar{c}(\tau)\bar{c}^{\dag}(0)\rangle=
 \langle T_{\tau} c(\tau)c^{\dag}(0)\rangle 
 \langle T_{\tau} X(\tau)X^{\dag}(0)\rangle$. 
A formula for the current can be derived using a Landauer-B\"uttiker 
approach.~\cite{meir92,jauho94} 
First we calculate the current from the left lead onto the dot from the 
rate of change of particles in the left lead. 
A similar expression for the current from the dot to the right lead is 
derived and 
the total current through the system is obtained 
by combining these two formulas. 
The derivation is presented in detail in Ref.~\onlinecite{meir92}
and \onlinecite{jauho94}. The result is that the current is given by 
\begin{equation}
I(V)=-\frac{2e}{h}\int d\epsilon \left[f_1(\epsilon)-f_3(\epsilon)\right] 
\mathrm{Im}\left[\mathrm{tr}(\Gamma G_2 (\epsilon) )\right]. 
\label{curr:eq}
\end{equation}
The applied voltage across the system is $V$ and it enters the two Fermi 
functions (the equilibrium Fermi level of the leads is chosen to be zero)
$f_1(\epsilon)=f(\epsilon-eV/2)$ and $f_3(\epsilon)=f(\epsilon+eV/2)$. 
Further, $G_2 (\epsilon) $ is the Green function for the quantum dot 
including all effects 
from the boson system {\it and} the tunneling
to the leads.
The parameter $\Gamma$ is 
\begin{equation}
\Gamma\equiv\frac{\Gamma_1\Gamma_3}{\Gamma_1+\Gamma_3},
\end{equation}
where $\Gamma_{1(3)}=2\pi t^2D_{1(3)}(\epsilon)$, $D_{1(3)}$ is the density 
of states (DOS) in the left (right) lead. $\Gamma_{1(3)}$ 
is the width of the central 
resonance due to the tunneling to the left ($\Gamma_1$) and right 
($\Gamma_3$) lead. The total width of the local resonance, $\Gamma_2$,  
is the sum of the two, $\Gamma_2=\Gamma_1+\Gamma_3$. 

For convenience we introduce the dimensionless parameters,
\begin{eqnarray}
g_1\equiv\left(\frac{M}{\hbar\omega_0}\right)^2 \nonumber \\
g_2\equiv\left(\frac{\Gamma}{\hbar\omega_0}\right)^2 \nonumber
\end{eqnarray}
We emphasize that there are many different energy scales
associated with the system:
$k_B T$, $e V$, $\hbar\omega_0$, $\Gamma$, $M$, and $\epsilon_0$.
The relative sizes of these energy scales have a significant
effect on the current through the system and what approximations
can be made in evaluating it.

The electrons will deposit/absorb energy from the bosonic system that 
has to be carried away/supplied. Therefore a question arises about how to 
define the temperature, particularly of the molecule or dot. 
We assume that the molecule/dot is 
in equilibrium with a bath and that the tunneling rate is small so that the 
system relaxes to the initial state after each tunneling event. 
In a quantum dot the bath can be the substrate that the quantum 
dot is manufactured on. For a molecule a surrounding cooling 
liquid~\cite{chen99} can play the role of the bath. Otherwise, we have 
to assume that the deposited or absorbed energy is transferred to/from the 
molecule via the leads. As far as we are aware, this assumption is also 
(implicitly) made in all other theoretical work on this subject. 

\section{Approximate evaluation of the Greens function $G_2(\epsilon) $}

To be able to use Eq.(\ref{curr:eq}) 
we have to calculate the local Green function, $G_2(\epsilon)$.
Due to the coupling to the leads finding 
$G_2 (\epsilon)$ is a highly non-trivial problem 
in many-body theory.~\cite{hewson80,hewson,haule} 
It is comparable in difficulty to the Kondo problem because
of the possibility of non-perturbative effects.
This is true even in equilibrium (i.e., in the absence of
a bias, $V=0$). A recent study was made of a similar
Hamiltonian (with spin) using the numerical renormalization
group.~\cite{hewson} We are interested in the non-equilibrium case
where there is a bias.
In order to simplify the analysis we have to rely on 
approximations, and the result will depend on how the $X$-operators 
from Eq.(\ref{X-op:eq}) are decoupled.  
One alternative is to assume that the coupling to the leads is 
small, $\Gamma_1+\Gamma_3 \ll \Delta$, this is the approach taken here. 
This approximation is justified for small currents, as is the case 
in the systems considered here. 
If we were to include the effect on the leads from the bosons on the 
molecule/dot there would be a narrowing effect on $\Gamma$. 
Hewson and Newns used variational and perturbation methods~\cite{hewson80}
to show that this narrowing only takes place if the following conditions apply:
\begin{eqnarray}
&&\frac{M^2}{\hbar\omega_0} > \Gamma, \ \ 
g_1 > 1, \ \ 
\hbar\omega_0 > \Gamma e^{-g_1}
  \nonumber\\
&&\hbar\omega_0 > |\Delta|.\nonumber 
\end{eqnarray}
The conditions on the first line means that the electron-boson coupling has 
to be large enough 
to form a polaron. The last requirement on the first line means that 
individual boson satellites can be distinguished from each other. 
The second line tells us that if the level and boson satellites are too far 
from the Fermi level it is energetically unlikely to have virtual boson
excitations, thus the leads are unaffected by the bosons. 
The narrowing is approximately given by
\begin{equation} 
t \rightarrow t e^{-g_1(1/2+n_B)},
\end{equation}
where $n_B$ is the Bose function,
\begin{equation}
n_B=\frac{1}{e^{\beta \hbar\omega_0}-1}
\end{equation}
and $\beta=1/k_BT$.

The above considerations apply to equilibrium ($V=0$)
whereas we are interested in the non-equilibrium situation
of a finite bias, and particularly the resonant tunneling
case where one of the leads' Fermi 
level is close to the dot/molecule level ($eV = \pm \Delta/2$).
In that case the narrowing of the level width due to that lead
(but not due to the second lead) may occur, e.g.,
\begin{equation}
\Gamma_2 = \Gamma_1 + \Gamma_3  \rightarrow \Gamma_1
+ \Gamma_3 e^{-g_1(1 + 2n_B)}.
\end{equation}
If $\Gamma_1 \sim \Gamma_3$ this will lead to some
quantitative but no significant qualitative changes
in the current-voltage characteristics and
so we will not consider them further. 

We treat the leads as unaffected by the bosons, i.e., no narrowing of the 
bands in the leads. This means that we ignore the averages of the 
$X$-operators that appear in the tunneling part of the 
Hamiltonian, Eq.(\ref{barHt:eq}), the justification for this 
is given above. Below we will also assume that 
the leads give rise to a flat, energy independent, density of states. This 
is sometimes called the wide band limit.~\cite{wingreen89} 
Otherwise $\Gamma$ would be energy dependent. The quantum dot Green 
function calculated using these approximations is 
\begin{equation}
G_2(t)=-i\Theta(t)e^{(i\Delta-\Gamma_2/2)t/\hbar}e^{-\Phi(t)}. 
\end{equation}
The factor $e^{-\Phi(t)}$ is due to the coupling to the boson and can 
be written~\cite{mahan} 
\begin{eqnarray}
&&\hspace*{-1cm}e^{-\Phi(t)}=
e^{-g_1(1+2n_B)} \nonumber \\
&&\times\sum_{l=-\infty}^{\infty}
I_l\left[2g_1\sqrt{n_B(1+n_B)}\right]e^{il\omega_0(t+i\beta/2)},
\label{expo:eq}
\end{eqnarray}
where $I_l$ denotes a modified Bessel function.

We Fourier transform 
the Green function and get an expression for the total current
\begin{eqnarray}
&&\hspace*{-1cm}I(V)=-\frac{e\Gamma}{h}\int_{-\infty}^{\infty} d\epsilon 
\left[f_1(\epsilon)-f_3(\epsilon)\right] 
e^{-g_1(1+2n_B)} \nonumber \\
&&\times \sum_{l=-\infty}^{\infty}
I_l\left[2g_1
         \sqrt{n_B(1+n_B)}\right]e^{-l\hbar\omega_0\beta/2} \nonumber \\ 
&&\times\frac{\Gamma_1+\Gamma_3}{(\epsilon+\Delta+l\hbar\omega_0)^2+
      \frac{(\Gamma_1+\Gamma_3)^2}{4}}.
\label{current:eq}
\end{eqnarray}

\begin{figure}[hb]
\includegraphics*[scale=1.0]{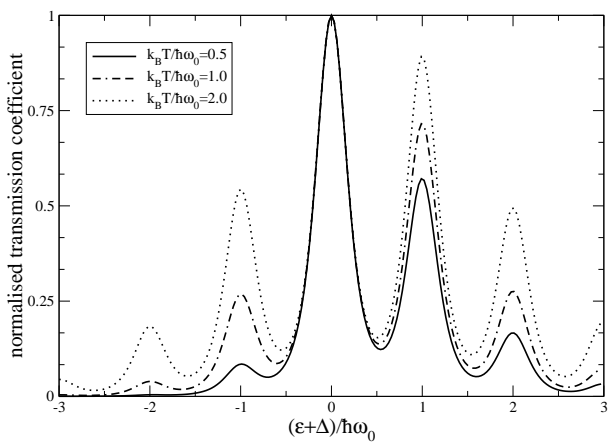}
\caption{Transmission coefficient, Im$\left[\mathrm{tr}(\Gamma G_2)\right]$, 
         as a function of the energy, for three
         different temperatures. 
         The satellites are due to the boson modes. 
         $g_1=\left(M/\hbar\omega_0\right)^2=0.5$ and 
         $g_2=\left(\Gamma/\hbar\omega_0\right)^2=0.09$. 
         The vertical axis is normalized to the highest peak in the plot.}
\label{trans:fig}
\end{figure}

We can interpret $\mathrm{Im}\left[\mathrm{tr}(\Gamma G_2)\right]$, in 
Eq.(\ref{curr:eq}), 
as the transmission coefficient for the tunneling. We plot this in 
Fig.~\ref{trans:fig} for a certain choice of parameters. 
The resonances to the left of $(\epsilon+\Delta)/\hbar\omega_0$ corresponds to 
absorption of bosons, and the ones to the right to emission of bosons. 
The middle line can be identified as the so called zero-boson transition. 
The width of each satellite depends on 
$\Gamma_2$ directly. When increasing the temperature the satellites 
increase in amplitude, indicating that it is easier to emit/absorb bosons. 
The asymmetry between negative and positive energies 
is due to the factor $e^{-l\hbar\omega_0\beta/2}$. 
This is a due to the fact that at low temperatures there are no available 
bosons to absorb. 

\begin{figure}[h]
\includegraphics*[scale=1.0]{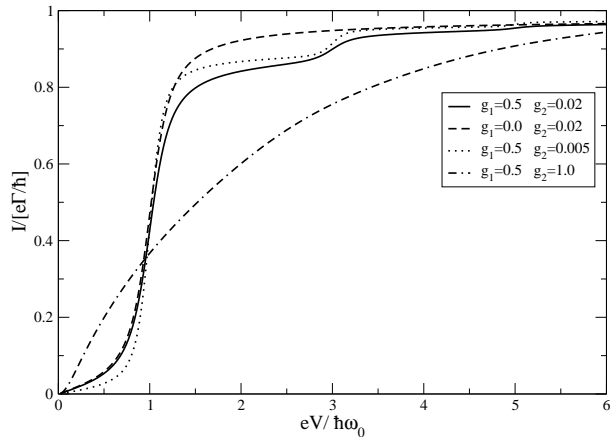}
\caption{Current as a function of the applied voltage for different 
         choice of coupling strengths.  We set
         $k_BT =0.1 \hbar\omega_0$ and $\Delta=\hbar\omega_0/2$.}
\label{I_V:fig}
\end{figure}

In Fig.~\ref{I_V:fig} we plot the current as a function of voltage using 
Eq.(\ref{current:eq}) for a set of parameters. 
It show steps 
indicating that more and more satellites participate in conducting electrons. 
Note that the steps in Fig.~\ref{I_V:fig} occur every second $\hbar\omega_0$. 
This is simply because the satellites are positioned equidistant on each side 
of the central resonance, we have to increase the voltage by $2\hbar\omega_0$ 
in order to cover the satellite.  The first satellite starts to 
contribute to the current when $eV=2\Delta$. 
A decrease of $\Gamma_2$ ($g_2$ decrease) results in sharper steps, and a 
decrease in the amplitude of the current. 
When $\Gamma \gg \hbar\omega_0$ (large $g_2$) the step structure disappears. 
Increasing the temperature results in 
the step structure being washed out to a smooth curve. 

\begin{figure}[h]
\includegraphics*[scale=1.0]{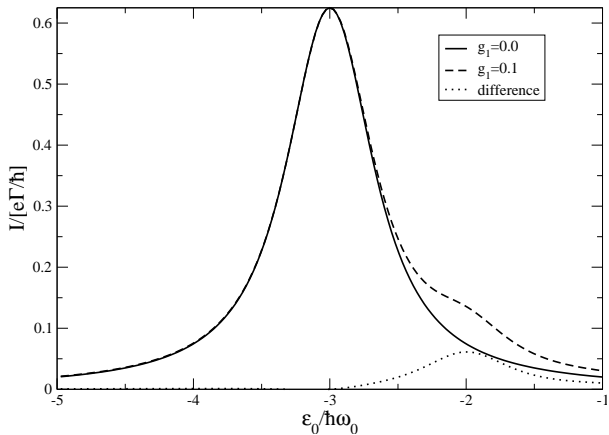}
\caption{Current as a function of the location
of the energy level in a quantum dot when bosons are 
         present ($g_1=0.1$) and absent ($g_1=0$). 
         $k_BT=0.03\hbar\omega_0$ and we put $g_2=0.5$. 
         $eV$ is set to $0.2\hbar\omega_0$ so that we only scan a small region 
         around $\epsilon_0$. 
         Parameters are taken from ref.~\onlinecite{qin01}. We only see the 
         boson emission satellite due to the low temperature and the small 
         electron-boson coupling.}
\label{IV_g:fig}
\end{figure}

When increasing $M$, 
the amplitude of the current drops due a decrease of the factor 
$e^{-\left(\frac{M}{\hbar\omega_0}\right)^2(1+2n_B)}$ 
in Eq.(\ref{current:eq}). Increasing the temperature has the same effect. 
Without any coupling to the boson ($M,g_1$=0) we get a single resonant level 
without any satellites. This can be seen in Fig.~\ref{IV_g:fig} where 
we plot the current as a function of $\epsilon_0$, the
location of the energy level in the dot
or molecule. The application 
of a gate voltage in a quantum dot would be equivalent to changing 
the level $\epsilon_0$ (or $\Delta$)~.\cite{fujisawa98,qin01} 
We see a shoulder developing 
corresponding to the first boson satellite. A 
similar effect has been seen in a 
double quantum dot system.~\cite{fujisawa98} 
The absence of a boson absorption peak in Fig.~\ref{IV_g:fig} is due to the 
low temperature, this comes from the factor 
$e^{l\hbar\omega_0\beta/2}$. If we increased the temperature, or the 
electron-boson coupling, enough there would be more side bands visible. 

\section{Limiting behavior for the current}

In order to better understand the influence from the bosons on the current. 
Let us now have a look at the current in some limits. 

\subsection{$M=0$}
If we put the coupling between the boson and the electrons to zero, 
we get: 
\begin{equation}
I(V)=-\frac{e}{h}\int_{-\infty}^{\infty} d\epsilon 
\left[ f_1(\epsilon)-f_3(\epsilon) \right]
      \frac{\Gamma_1\Gamma_3}{(\epsilon-\epsilon_0)^2+
                              \frac{(\Gamma_1+\Gamma_3)^2}{4}}.
\label{Im0}
\end{equation}
This would correspond to resonant tunneling without any bosons. 

\subsubsection{$ k_B T \gg \Gamma, eV$}   

In this limit Eq.(\ref{Im0}) reduces to the linear response expression
\begin{equation}
\frac{I}{V}
=\frac{4e^2}{h} \frac{\pi\Gamma}
{k_BT\cosh^2 \left(\frac{\epsilon_0}{k_B T}\right)}
\end{equation}
A similar form was used by Qin {\it et.\ al}~\cite{qin01}
to fit their experimental data.

\subsubsection{$T=0$}
If the temperature goes to zero we can approximate the 
Fermi functions with step functions. Then, the integral over $\epsilon$ can 
be performed and the result is
\begin{equation}
\lim_{T\rightarrow 0}I(V)=\frac{2e\Gamma}{h} 
\left[ \tan^{-1}\left(\frac{eV-2\epsilon_0}{\Gamma_1+\Gamma_3}\right) 
 +\tan^{-1}\left(\frac{eV+2\epsilon_0}{\Gamma_1+\Gamma_3}\right)\right]. 
\label{limit:eq}
\end{equation}
Further, if $eV$ and $2\epsilon_0$ is small compared to $\Gamma_1+\Gamma_3$ we 
can use the property that $\tan^{-1}(x)\sim x$, and we get 
\begin{equation}
\lim_{\stackrel{T\rightarrow 0}{(eV,2\epsilon_0)/(\Gamma_1+\Gamma_3)\ll 1}}I=
\frac{4e^2\Gamma}{h(\Gamma_1+\Gamma_3)}V, 
\end{equation}
i.e., a \textit{linear} regime at low voltages. 
If, on the other hand, we take $V\rightarrow\infty$ in Eq.(\ref{limit:eq}) 
we get 
\begin{equation}
\lim_{\stackrel{T\rightarrow 0}{V\rightarrow \infty}}I=
\frac{e\Gamma}{\hbar}. 
\label{lim_T_V:eq}
\end{equation}
This means that the whole resonant level contributes maximal to the current. 

\subsection{$M\neq 0$}

\subsubsection{$eV \gg k_BT,\hbar\omega_0$}

In this case we get the same limit as in Eq.(\ref{lim_T_V:eq}) 
even if $M\neq0$ from Eq.(\ref{current:eq}). 
This can be seen in Fig.~\ref{I_V:fig} 
where all curves tend to the same value at large $V$. 
If we have that $eV\gg \Delta, k_BT$ we can replace 
$f_1(\epsilon)-f_3(\epsilon)$ by a factor 1, and the integral would extend 
between $-eV/2$ and $eV/2$. But since $eV$ is greater than all other energies 
we extend the integral from $-\infty$ to $\infty$. The integral gives a 
contribution $\pi$. All parts coming from the boson gives 1 and we again 
have the limit
\begin{equation}
I(eV\gg \Delta, k_BT) \simeq \frac{e\Gamma}{\hbar}. 
\end{equation}
This limit can be seen in Fig.~\ref{I_V:fig} where all curves tend to the 
same limit at high applied voltage. 

\subsubsection{$k_BT \ll \hbar\omega_0$}

Let us now investigate the limit $k_BT \ll \hbar\omega_0$.
In this limit (corresponding to low temperatures) we can approximate the Bose 
function as $n_B\simeq e^{-\hbar\omega_0/k_BT}\ll 1$. 
All terms corresponding to positive $l$ vanishes. 
This is a result of the physical fact that positive $l$ corresponds to  
boson absorption but at $T=0$ there are no bosons. 
The Bessel function can be approximated as
$I_l(z)\sim\frac{1}{l!}(z/2)^2$ when $z\rightarrow 0$. 
Then we get that the current becomes 
\begin{eqnarray}
&&\hspace*{-1cm}I_{k_BT \ll \hbar\omega_0}=-
\frac{2e\Gamma}{h} e^{-g_1}\int_{-\infty}^{\infty} d\epsilon 
\left[f_1(\epsilon)-f_3(\epsilon)\right] \nonumber \\
&&\times\sum_{l=-\infty}^{0}\frac{g_1^{|l|}}{|l|!}
      \frac{\frac{\Gamma_1+\Gamma_3}{2}}{(\epsilon+\Delta+l\hbar\omega_0)^2+
      \frac{(\Gamma_1+\Gamma_3)^2}{4}}.
\end{eqnarray}

\subsubsection{$k_BT \gg \hbar\omega_0$}

For high temperatures we approximate the 
Bose function as $n_B\simeq k_BT/\hbar\omega_0$. The argument 
in the Bessel function is large and we can use the property 
\begin{equation}
I_l(z) \simeq \frac{e^z}{\sqrt{2\pi z}},\;\;  z \gg 1.
\end{equation}
Using this, the current becomes
\begin{eqnarray}
&&\hspace*{-1cm}I_{k_BT\gg\hbar\omega_0}=-
\frac{2e\Gamma}{h}
\frac{e^{-g_1\frac{\hbar\omega_0}{4k_BT}}}
{\sqrt{4\pi g_1k_BT/\hbar\omega_0}}
\int_{-\infty}^{\infty} d\epsilon
\left[f_1(\epsilon)-f_3(\epsilon)\right] \nonumber \\
&&\times\sum_{l=-\infty}^{\infty}
        \frac{\frac{\Gamma_1+\Gamma_3}{2}}{(\epsilon+\Delta+l\hbar\omega_0)^2+
      \frac{(\Gamma_1+\Gamma_3)^2}{4}}.
\label{limit2:eq}
\end{eqnarray}

\subsection{Saddle point approximation}

If $g_2\gg 1$ (i.e., $\Gamma \gg \hbar \omega_0$)
we can evaluate the current using a saddle-point 
approximation similar to
that used previously in Ref.~\onlinecite{shotte66} and \onlinecite{armour01}. 
The exponential factor, $e^{-\Phi(t)}\equiv \langle X(t) X^{\dag}(0) \rangle$, 
in Eq.(\ref{expo:eq}) can be written as 
\begin{equation}
e^{-g_1[(n_B+1)(1-e^{-i\omega_0t})+n_B(1-e^{i\omega_0t})]}. 
\end{equation}
We approximate the exponential 
function in the exponent, $e^{z}\sim 1+z+z^2/2$, 
and we get:
\begin{equation}
G_2(t)\simeq -ie^{i(\Delta+g_1\hbar\omega_0)t/\hbar-\Gamma_2t/2\hbar-g_1/2
                  (1+2n_B)(\omega_0t)^2}.
\end{equation}
Let us assume that we can neglect the term 
linear in $t$ in the exponent compared to 
the quadratic one, i.e., $g_1 \gg g_2$. 
We Fourier transform the resulting Green function and get that the
relevant  factor entering Eq.(\ref{curr:eq}) becomes
\begin{equation}
\mathrm{Im}\left[G_2(\epsilon)\right]\simeq 
\frac{\exp\left[-\frac{(g_1\hbar\omega_0+\Delta+\epsilon)^2}
                      {2g_1(\hbar\omega_0)^2(1+2n_B)}\right]}
     {\omega_0\sqrt{g_1(1+2n_B)}}.
\label{saddle2:eq}
\end{equation}
This approximation gives a broad
Gaussian line shape covering all the boson satellites.
This is in contrast to the 
individual boson satellites shown in Fig.~\ref{trans:fig}.
Using the saddle point approximation would give a Gaussian 
line shape in $I(\epsilon_0)$,
whereas a Lorentzian line shape occurs in the regime, 
$k_BT \ll \Gamma \ll \hbar\omega_0$, illustrated in Fig.~\ref{IV_g:fig}.

\begin{figure}[h]
\includegraphics*[scale=1.0]{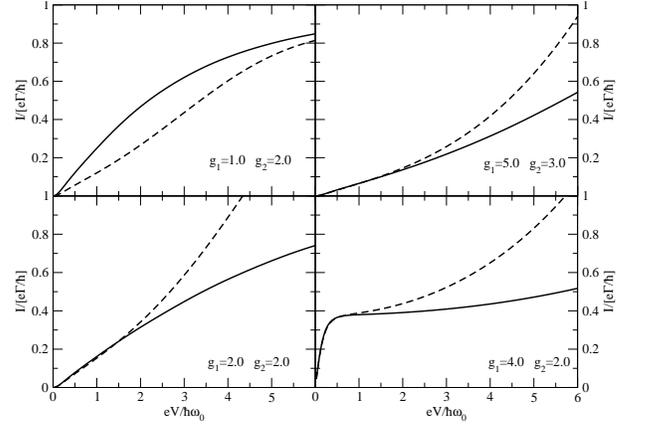}
\caption{Failure of the saddle point approximation. 
         Current calculated in two different ways: the full lines were 
         obtained using the exact result (Eq.(\ref{current:eq}))
         and the dashed lines using the 
	 saddle point approximation (Eq.(\ref{saddle2:eq})) 
         in the expression for the current. We see that the 
         saddle approximation does not reproduce the full expression for 
         the current. 
 Here we set
         $k_BT = 0.1 \hbar\omega_0$ and $\Delta=\hbar\omega_0/2$.}
\label{comparison:fig}
\end{figure}

The current using Eq.(\ref{saddle2:eq}) is plotted in 
Fig.~\ref{comparison:fig}, and compared to the full
expression, Eq.(\ref{current:eq}).
In this figure we can clearly see that the saddle point approximation cannot 
reproduce the actual current. Only for a 
small range of bias voltages, for low 
temperature and large coupling is there an agreement. 

\section{Differential conductance}

The differential conductance, defined by   
\begin{equation}
C=\frac{dI}{dV},
\end{equation}
more clearly reveals the effect of the bosons.
In general this is given by
\begin{eqnarray}
&&\hspace*{-1cm}C =\frac{e^2\Gamma\beta}{h}
\int_{-\infty}^{\infty} d\epsilon 
\left\{f_1(\epsilon)\left[1-f_1(\epsilon)\right] +
       f_3(\epsilon)\left[1-f_3(\epsilon)\right] \right\}\nonumber \\
&&\times e^{-g_1(1+2n_B)} 
\sum_{l=-\infty}^{\infty}
I_l\left[2g_1\sqrt{n_B(1+n_B)}\right] \nonumber \\ 
&&\times e^{-l\hbar\omega_0\beta/2} 
\frac{\frac{\Gamma_1+\Gamma_3}{2}}{(\epsilon+\Delta+l\hbar\omega_0)^2+
      \frac{(\Gamma_1+\Gamma_3)^2}{4}}.
\label{cond:eq}
\end{eqnarray}
Later we will set $eV=2\Delta$, which corresponds to resonant transport
through the zero phonon feature.
If we let the temperature go to zero in this expression 
we can approximate the Fermi functions together with the temperature 
as a delta function, 
$\beta n_f(\epsilon)[1-n_f(\epsilon)]\sim \delta(\epsilon)$, 
and again only negative $l$ contributes, corresponding to emission of bosons, 
and we get
\begin{eqnarray}
&&\hspace*{-1cm}[C(V)]_{T\rightarrow 0}=
  \frac{e^2\Gamma_1\Gamma_3}{2h}
     e^{-g_1} \sum_{l=-\infty}^{0}\frac{g_1^{|l|}}{|l|!}
\nonumber \\
&&\times\left[
     \frac{1}{(\Delta+l\hbar\omega_0+eV/2)^2+\frac{(\Gamma_1+\Gamma_3)^2}{4}}
\right.\nonumber \\
&&\left.
    +\frac{1}{(\Delta+l\hbar\omega_0-eV/2)^2+\frac{(\Gamma_1+\Gamma_3)^2}{4}}
     \right]. 
\end{eqnarray}
We define
\begin{equation}
(C_{\mathrm{res}})^0\equiv [C(eV=2\Delta)]_{T\rightarrow 0}.
\end{equation}

For the particular case $\Delta \gg \hbar \omega_0, g_1 \hbar \omega_0$,
this simplifies to
$(C_{\mathrm{res}})^0 \simeq \frac{2 e^2}{h}
 \frac{\Gamma}{\Gamma_1 + \Gamma_3} e^{-g_1}$, showing
how polaronic effects reduce the differential conductance.

\begin{figure}[hb]
\includegraphics*[scale=1.0]{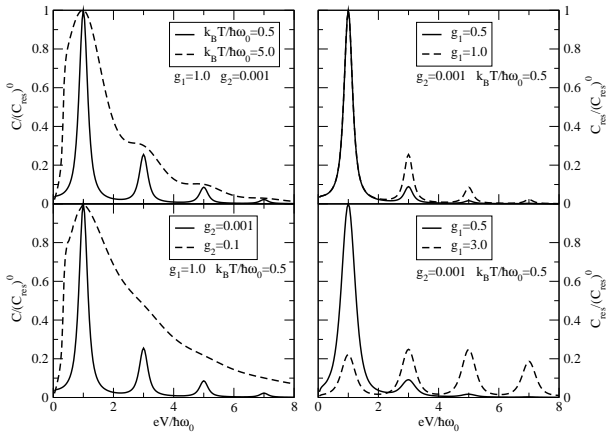}
\caption{Differential conductance as a function of applied voltage 
         when changing the temperature (upper left), electron-boson 
         coupling $g_1$ (upper right and lower right) and the level 
         width $g_2$ (lower left). $\Delta=\hbar\omega_0/2$.
         At $eV=2\Delta$ it has a maxima for moderate couplings 
         $g_1\alt 1$. To obtain a maximal signal 
         it would be desirable to perform the experiments at this value.}
\label{diff_cond:fig}
\end{figure}

In Fig.~\ref{diff_cond:fig} we plot the differential conductance as a 
function of the applied voltage for different values of temperature and 
coupling parameters. 
The peak at $eV=2\Delta$ correspond to the zero-boson peak, and in the 
consecutive peaks one/two/three$\ldots$, and so on, bosons are emitted 
or absorbed.
As seen in this figure increasing the temperature, or the level widths, 
drastically affects the shape of the differential conductance. 

\begin{figure}[h]
\includegraphics*[scale=1.0]{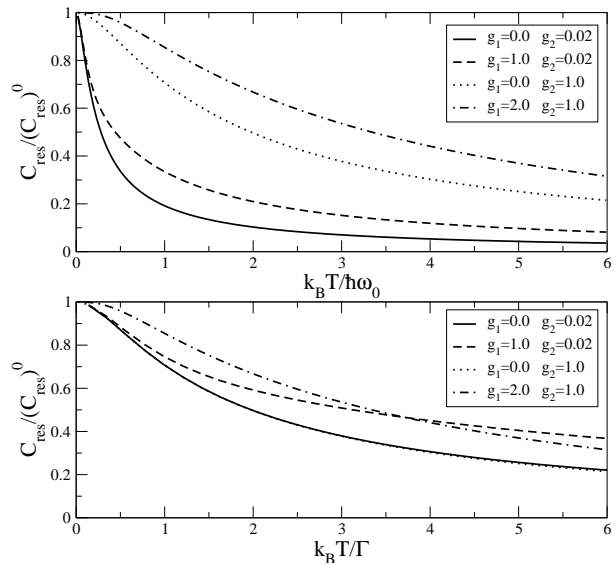}
\caption{Differential conductance at the resonance as a function of 
         temperature for different parameters. 
         The upper graph shows that, for moderate couplings 
         $g_1\alt 1$, the differential conductance is 
         almost unaltered by the presence of the bosons. 
         The lower graph shows that the 
         differential conductance, for moderate couplings $g_1\alt 1$,
         is determined by the parameter $k_BT/\Gamma$. 
         The plots were made assuming a constant DOS in the leads, 
         $\Delta=0$ and $eV=0$.}
\label{s_T:fig}
\end{figure}

In Fig.~\ref{s_T:fig} we plot the 
differential conductance on resonance 
with the zero phonon line as a function of temperature for a range
of parameters. 
In  this figure we see that the differential conductance generally
decreases with increasing temperature,
in contrast to the non-monotonic dependence found
by Holstein~\cite{holstein59} for periodic
molecular crystals.
The corresponding  crossover behavior \textit{does not occur}
for transport through molecules/quantum dots, 
since this would be indicated by an upturn in Fig.~\ref{s_T:fig} when 
increasing the temperature. The absence of a crossover 
can also be seen by looking at Fig.~\ref{I_V:fig} from that the slope 
at $eV=2\Delta$ (i.e., the differential conductance) is almost constant 
when changing $g_1$ from 0 to 0.5. If we were to calculate the differential 
conductance  
in the limit when $k_BT\gg\hbar\omega_0$ from Eq.(\ref{limit2:eq}) we see 
that the temperature dependence of the differential conductance  is governed 
by the pre-factor 
$\frac{e^{-g_1\frac{\hbar\omega_0}{4k_BT}}}
         {\sqrt{4\pi g_1k_BT/\hbar\omega_0}}$ 
and this is a strictly \textit{decreasing} function of the temperature, 
for reasonably values of $g_1$. 
Thus, there will never be an upturn in the differential conductance  when 
increasing the temperature. This general behavior is not changed when 
$g_1$ is changed. Even an increased applied voltage, meaning that more 
boson satellites contributes, was not able to induce a crossover. 
However, changing $\Gamma$ does alter the amplitude 
of the differential conductance, as seen in Fig.~\ref{s_T:fig}. 

As mentioned above the temperature behavior is dominated by $\Gamma$. 
If we put $M=0$ in Eq.(\ref{cond:eq}) we can write the differential 
conductance as
\begin{equation}
[C_{res}]_{M\rightarrow0}=
\frac{e^2\Gamma}{h}\frac{\tilde{\Gamma}}{k_BT}\int_{-\infty}^{\infty}
dy[f'(y)+f'(y+\frac{2\epsilon_0}{k_BT})]\frac{1}{y^2+(\tilde{\Gamma}/k_BT)^2},
\end{equation}
where $\tilde{\Gamma}\equiv(\Gamma_1+\Gamma_3)/2$. 
If we now take $\epsilon_0=0$ 
or $k_B T \ll \epsilon_0$ we will have that the differential 
conductance  is a \textit{universal} function of $\tilde{\Gamma}/k_BT$, 
i.e,
\begin{equation}
[C_{res}]_{M\rightarrow0}=F(\tilde{\Gamma}/k_BT).
\end{equation}
This can be seen in the lower graph in Fig.~\ref{s_T:fig}, where the two 
graphs for $g_1=0$ (but $g_2=0.02$ and $2.0$ respectively) collapse 
on the same line when the temperature axis is rescaled. 

\section{Conclusion}
In conclusion we see that the polaronic transport
through a single molecule or quantum dot
does not clearly exhibit the crossover from coherent to 
incoherent transport expected for  the  case of 
a periodic molecular crystal considered by 
Holstein~\cite{holstein59}. The general behavior of the 
temperature dependence of the 
differential conductance  is in large \textit{unaffected} by the 
presence of the bosons. 
The temperature dependence is mostly determined by 
the linewidth (due to coupling to the leads)
of the resonant energy level. The bosons produce side bands corresponding to 
absorption and emission of bosons. 
We also stressed that because of the interaction of the polaron
on the dot or molecule with the leads there are potentially
some very interesting problems in many-body physics~\cite{hewson80,hewson} to 
be explored in the model system we have considered.

U.\ Lundin acknowledges the support from the Swedish Foundation for
International Cooperation in Research and Higher Education (STINT).
This work was also supported by the 
Australian Research Council (ARC). 

\bibliography{paper}

\end{document}